\documentclass[12pt]{iopart}
\usepackage{iopams}  
\usepackage[applemac]{inputenc}

\def\ceil#1{\lceil #1 \rceil}
\begin{document}
\title{Infinite Shannon entropy}
\author{Valentina Baccetti and Matt Visser}
\address{School of Mathematics, Statistics, and Operations Research\\
Victoria University of Wellington, PO Box 600, \\
Wellington 6140, New Zealand}
\ead{valentina.baccetti@msor.vuw.ac.nz and matt.visser@msor.vuw.ac.nz}
\begin{abstract}
Even if a probability distribution is properly normalizable, its associated Shannon (or von Neumann) entropy can easily be infinite. 
We carefully analyze conditions under which this phenomenon can occur. 
Roughly speaking, this happens when arbitrarily small amounts of probability are dispersed into an infinite number of states; we shall quantify this observation and make it precise. We develop several particularly simple, elementary, and useful bounds,  and also provide some asymptotic estimates, leading to necessary and sufficient conditions for the occurrence of infinite Shannon entropy. We go to some effort to keep technical computations as simple and conceptually clear as possible. In particular, we shall see that large entropies cannot be localized in state space; large entropies can only be supported on an exponentially large number of states.
We are for the time being interested in single-channel Shannon entropy in the information theoretic sense, not entropy in a stochastic field theory or QFT defined over some configuration space, on the grounds that this simple problem is a necessary precursor to understanding infinite entropy in a field theoretic context.


\vskip 10 pt
\noindent 
21 December 2012; 7 January 2013; \LaTeX-ed \today

\end{abstract}


\maketitle

\bigskip
\hrule
\bigskip
\markboth{Infinite Shannon entropy}{}
\tableofcontents
\markboth{Infinite Shannon entropy}{}
\bigskip
\hrule
\bigskip
\clearpage
\markboth{Infinite Shannon entropy}{}
\def\d{{\mathrm{d}}}
\section{Introduction}\label{S:intro}
The classical Shannon entropy, (and closely related quantum von Neumann entropy), is a general purpose theoretical tool with a vast number of applications ranging from engineering to demographics to quantum information theory and other branches of theoretical physics~\cite{Shannon:1948, Shannon:1949, Ash:1965, Yeung:2002, Watrous:2008, Renner:2009}. Stripped to its essentials, one considers a set of normalized probabilities $\sum_n p_n = 1$, and analyzes the properties of the quantity:
\begin{equation}
S = - \sum_n p_n \ln p_n.
\end{equation}
Two major cases are of interest, when the index set $\{n\}$ characterizing the various ``states’’ of the system is finite, and when it is countably infinite. A third case, when the  index set $\{n\}$  is uncountably infinite, requires an integral formulation of the entropy, and we shall not presently have anything specific to say about this uncountable case. One way of justifying such a restriction is via an appeal to quantum mechanics where, in terms of a normalized density matrix $\tr[\rho]=1$, the von Neumann entropy is:
\begin{equation}
S = - \tr[ \rho \ln \rho].
\end{equation}
If, (as is usual), quantum physics is formulated on a \emph{separable} Hilbert space, then the density matrix can be diagonalized over a countable basis, and the von Neumann entropy reduces to the Shannon entropy over a countable (or possibly even finite) set of states.  For this reason we shall restrict attention to the finite or countably infinite cases. 

If the total number of states is finite, $N = \sum_n 1 <\infty$, then an elementary computation leads to the upper bound
\begin{equation}
S \leq \ln N. 
\end{equation}
More subtly, even if the total number of states is infinite, $N = \sum_n 1 = \infty$, then as long the total number of states of non-zero probability is finite, $N’ = \sum_{n: p_n > 0} 1 <\infty$, an equally elementary computation leads to the upper bound
\begin{equation}
S \leq \ln N’. 
\end{equation}
These simple observations demonstrate that to obtain infinite Shannon entropy, an infinite number of states must have non-zero probability, in particular:
\begin{equation}
N’ \geq \exp S. 
\end{equation}
But then, since the sum of the probabilities is unity, an infinite number of states must have \emph{non-zero but arbitrarily small} probability. We shall now seek to quantify these observations in a straightforward and transparent manner.  Some earlier rather technical work along these lines can be found in~\cite{finite}.  Our own interest in these issues was prompted by the more general issues raised in~\cite{Wolf, area, Wehrl:1978, Wehrl:1976, Eisert:2002, bounds, Bekenstein:1973, Bombelli:1986, Srednicki:1993, Jacobson:1995}.  In the long run we are interested in infinite entropies that can arise in stochastic field theories and QFTs, but the single channel information theoretic context of the current article already provides some interesting subtleties.

\section{Some examples of infinite Shannon entropy}\label{S:infinite}

To show that situations of infinite Shannon entropy can indeed occur perhaps the simplest example is to consider the sum:
\begin{equation}
\Sigma(u) = \sum_{n=\ceil{e}}^\infty  {1\over n \; (\ln n)^{1+u}}; \quad \hbox{(converges for $u>0$, diverges for $u\leq 0$)}. 
\end{equation}
Variants of this series are discussed for instance in Hardy~\cite{pure}, Hardy and Riesz~\cite[see esp p 5]{dirichlet-series}, and Shilov~\cite[see esp \S 6.15.c on p 192]{shilov}. Here $\ceil{x}$ is the ``ceiling’’ function, the smallest integer $\geq x$. The perhaps somewhat unexpected lower limit of summation $\ceil{e}$ is designed to ensure that $\ln n> 0$ for all terms in the sum, so that one never has to raise a negative number to a real power. 
The corresponding probabilities (defined only for $n\geq \ceil{e}$) are
\begin{equation}
p_n =  {1\over \Sigma(u) \; n \; (\ln n)^{1+u}}.
\end{equation}
These are well defined and properly normalized for $u>0$. But then
\begin{eqnarray}
S &=& \sum_n  {1\over \Sigma(u)  \;n \; (\ln n)^{1+u}} \ln\{  \Sigma(u)  \;n \;(\ln n)^{1+u} \} 
\\
&=& \ln \Sigma(u) +  {1\over \Sigma(u) } \sum_n  {1\over n\; (\ln n)^{u}} + {1+u\over\Sigma(u)} \sum_n  {\ln\ln n\over n \; (\ln n)^{1+u}}
\\
&= & \ln \Sigma(u) +  {1\over \Sigma(u) } \sum_n  {1\over n \; (\ln n)^{u}} - {\d\Sigma(u)/\d u\over\Sigma(u)}.
\end{eqnarray}
The first and third terms converge for $u>0$, but the second term converges only for $u>1$. So this probability distribution is convergent but has infinite Shannon entropy over the entire range $u\in(0,1]$. 
The particular case $n=1$, where $p_n \propto {1\over n (\ln n)^2}$ has previously been considered in a number of specific situations~\cite{Wehrl:1978,  Bender:2005}.
Apart from the entire range $u\in (0,1]$ above, there are many more examples along similar lines. For instance one could consider the sums
\begin{equation}
\Sigma_2(u) = \sum_{n=\ceil{e^e}}^\infty  {1\over n \;\ln n\; (\ln\ln n)^{1+u}}; 
\end{equation}
\begin{equation}
\Sigma_3(u) = \sum_{n=\ceil{e^{e^e}}}^\infty  {1\over n \;\ln n\; \ln\ln n\;(\ln\ln\ln n)^{1+u}}; 
\end{equation}
both of which converge for $u>0$ and diverge for $u\leq 0$,
and the obvious infinite chain of generalizations thereof.  Thereby (following the analysis above) one can easily construct an infinite chain of probability distributions that are convergent (and so are properly normalized) for $u>0$ but whose Shannon entropy converges only for $u>1$. These probability distributions are all convergent but have infinite Shannon entropy over the entire range $u\in(0,1]$. 
Even more baroque examples are possible, (but perhaps not desirable).

\section{Probability gap}\label{S:gap}

As a first step towards analyzing and quantifying the conditions under which infinite Shannon entropy can occur, let us define a notion of ``probability gap’’  when there is a minimum non-zero probability. (The idea is to mimic the notion of ``mass gap’’.) More precisely, let
\begin{equation}
p_* = \inf\{ p_n: p_n > 0\}.
\end{equation}
If $p_*>0$ then, (since $1= \sum_n p_n \geq N’ \; p_*$),  we have $N’\leq 1/p_* \leq \infty$. We see that only a finite number of the $p_n$ can then be non-zero, and in fact the \emph{infimum} can be replaced by a \emph{minimum}:
\begin{equation}
p_* = \min\{ p_n: p_n > 0\}.
\end{equation}
So for all practical purposes the presence of a probability gap means the state space is effectively finite. Conversely if only a finite number of probabilities are non-zero then there is a probability gap. In particular it is elementary that
\begin{equation}
S = - \sum_n p_n \ln p_n \leq - \sum_n p_n \ln p_* = - \ln p_*  < \infty, 
\end{equation}
though a slightly stronger result is also available
\begin{equation}
S \leq \ln N’ \leq -\ln p_*  < \infty.
\end{equation}
So we see very explicitly that for infinite Shannon entropy one cannot have a probability gap.

\section{Elementary bound leading to the Gibbs inequality}\label{S:gibbs}
Let us now try to be more quantitative. 
Based on the utterly elementary fact that for positive $x$ we have $[x\ln x]’{}’ = 1/x> 0$, it is immediate that for positive numbers
\begin{equation}
x \ln(x/a) + y\ln(y/b) \geq (x+y) \ln\left({x+y\over a+b}\right),
\end{equation}
with equality only when $x/a=y/b$. 
See \cite[p 97 \S 117]{inequalities}.\\
Proof: Since the second derivative is positive
\begin{equation}
{a\over a+b} \; \tilde x \ln  \tilde x + {b\over a+b} \;  \tilde y \ln y \geq {a  \tilde x + b \tilde y\over a+b} \ln\left( {a  \tilde x + b \tilde  y\over a+b} \right),
\end{equation}
with equality only when $ \tilde  x = \tilde y$. 
Therefore
\begin{equation}
a  \tilde x \ln  \tilde x + b  \tilde y \ln  \tilde y \geq (a  \tilde x + b \tilde y) \ln\left( {a  \tilde x + b \tilde y\over a+b} \right).
\end{equation}
Now rename $a \tilde x\to x$ and $b \tilde y \to y$ to obtain the desired result. \hfill $\Box$
\\
(It is worth explicitly verifying this since the justification is so elementary, and the payoff will be immediate.) 
Now iterate this result:
\begin{equation}
x \ln(x/a) + y\ln(y/b) +z \ln(z/c) \geq (x+y+z) \ln\left({x+y+z\over a+b+c}\right).
\end{equation}
More generally, for positive $x_n$ and $a_n$:
\begin{equation}
\sum_n x_n \ln(x_n/a_n) \geq \left(\sum_n x_n\right) \ln\left({\sum_n x_n\over \sum_n a_n}\right).
\end{equation}
When $\sum_n p_n = \sum_n q_n = 1$ the above gives an elementary proof that
\begin{equation}
\sum_n p_n \ln(p_n/q_n) \geq \left(\sum_n p_n\right) \ln\left({\sum_n p_n\over \sum q_n}\right) = 0.
\end{equation}
That is
\begin{equation}
\sum_n p_n \ln p_n \geq \sum_n p_n \ln q_n ,
\end{equation}
so
\begin{equation}
S \leq - \sum_n p_n \ln q_n.
\label{E:gibbs}
\end{equation}
This result is of course extremely well known, typically being referred to as the Gibbs inequality, (or the positivity theorem for relative entropy), with proofs most typically involving a less than elementary appeal to Jensen’s inequality. We shall now apply this result to the matter at hand.

\section{Partial counts, partial probabilities, partial Shannon entropies}\label{S:partial}

Let us now consider the effect of summing only over some restricted subset $X$ of the total state space $\{n\}$. Define
\begin{equation}
N_X = \sum_{n\in X} 1; \qquad P_X = \sum_{n\in X} p_n; \qquad S_X = - \sum_{n\in X}  p_n \ln p_n.
\end{equation}
In particular, using the inequality demonstrated above, we have
\begin{equation}
\sum_{n\in X} p_n \ln p_n \geq \left(\sum_{n\in X}  p_n\right) \ln\left({\sum_{n\in X}  p_n\over \sum_{n\in X}  1}\right) = P_X \ln(P_X/N_X).
\end{equation}
Therefore
\begin{equation}
S_X \leq P_X [\ln N_X - \ln P_X].
\label{E:ineq-a}
\end{equation}
Though this looks very similar to the entropy bound derived for the total entropy over a finite state space, there are significant differences --- the current bound now tells you something deeper about the extent to which entropy can be localized in the state space. Indeed we can recast the bound as:
\begin{equation}
N_X \geq P_X \; \exp(S_X/P_X).
\label{E:ineq-a2}
\end{equation}
That is, packing a finite amount of entropy $S_X$ into a region containing total probability $P_X$ requires an exponentially large number of states $N_X$.

Another way of interpreting this bound is to define the average probability per state, and average entropy per state,  over the set $X$ by:
\begin{equation}
\bar P_X = {P_X\over N_X}; \qquad \bar S_X = {S_X\over N_X}.
\end{equation}
Then
\begin{equation}
\bar S_X \leq - \bar P_X \ln \bar P_X.
\end{equation}
A slightly weaker but perhaps more intuitive bound is obtained if we vary the RHS of  equation (\ref{E:ineq-a}) with respect to $P_X$ while holding $N_X$ fixed (assume $N_X\geq \ceil{e}=3$). Then
\begin{equation}
{\partial \,\hbox{RHS}\over\partial P_X} = \ln(N_X/P_X) - 1 > 0.
\end{equation}
So the maximum of the RHS occurs for $P_X=1$, and we see (for $N_X\geq 3$)
\begin{equation}
S_X < \ln N_X. 
\label{E:ineq-b}
\end{equation}
%
%
Similarly if we vary the RHS of equation (\ref{E:ineq-a2}) with respect to $P_X$ while holding $S_X$ fixed (assume $S_X\geq1$). Then
\begin{equation}
{\partial \,\hbox{RHS}\over\partial P_X} = \exp(S_X/P_X) \left\{1 - S_X/P_X\right\} < 0.
\end{equation}
So the minimum of the RHS occurs for $P_X=1$, and we see (for $S_X\geq 1$)
\begin{equation}
N_X > \exp S_X. 
\label{E:ineq-b2}
\end{equation}
The message to take from the logarithmic and exponential bounds is again that large Shannon entropies cannot be tightly localized in state space, large Shannon entropies must invariably come from exponentially large ($N_X > \exp S_X$) regions of state space.

\section{Asymptotic estimates}\label{S:asymptotic}

Let us now consider the effect of adding some extra order-based structure, by summing only over the high-probability sector of the total state space $\{n\}$. Define the quantities:
\begin{equation}
N(p) = \sum_{n: p_n \geq p} 1; \qquad P(p) = \sum_{n: p_n \geq p}  p_n; \qquad S(p) = - \sum_{n: p_n \geq p}   p_n \ln p_n.
\end{equation}
These are ``probability cutoff’’ sums where the low probability events are excluded.
Note 
\begin{equation}
\lim_{p\to0} N(p) \to N; \qquad \lim_{p\to0} P(p) \to 1; \qquad \lim_{p\to0} S(p) \to S;
\end{equation}
where $N$ and $S$  may be infinite. 
It may sometimes be useful to define $N’ = \lim_{p\to0^+} N(p)$ (which may again be infinite) in order to explicitly exclude the zero modes.
Now
\begin{equation}
S(p) \leq \sum_{n: p_n \geq p} p_n [-\ln p] = - P(p) \ln p \leq - \ln p.
\end{equation}
That is
\begin{equation}
S(p)  \leq - \ln p,
\end{equation}
so in some sense the total Shannon entropy can never be worse than ``logarithmically divergent’’ in the probability cutoff. 
Similarly
\begin{equation}
1 \geq P(p) \geq p N(p); \qquad \hbox{that is} \qquad N(p) \leq {P(p) \over p}  \leq {1 \over p}.
\end{equation}
We also have
\begin{equation}
S(p) \leq P(p) [\ln N(p) - \ln P(p)].
\label{E:ineq-aa}
\end{equation}
Combining these results we regain
\begin{equation}
S(p) \leq - P(p) \ln p \leq - \ln p.
\label{E:ineq-aaa}
\end{equation}
We again see that to get infinite Shannon entropy one needs an infinitely large number of arbitrarily low probability events. 

\section{Entropy bounds from the Gibbs inequality}\label{S:gibbs-bounds}

Let us now obtain several explicit bounds directly from the Gibbs inequality.
Consider the quantities $q_n = n^{-z}/\zeta(z)$ where $\zeta(z)$ is the Riemann zeta function. Then we have $\sum_{n=1}^\infty q_n=1$ for $z>1$.  The Gibbs inequality becomes
\begin{equation}
S \leq - \sum_{n=1}^\infty p_n\, \ln q_n = \ln \zeta(z) + z \sum_{n=1}^\infty p_n \ln n.
\label{E:gibbs1}
\end{equation}
Thus a \emph{sufficient} condition for the Shannon entropy to be finite is 
\begin{equation}
\langle \ln n \rangle = \sum_{n=1}^\infty p_n \, \ln n < \infty.
\end{equation}
A number of quite similar results can easily be obtained:
\begin{itemize}
\item 
Consider for instance the quantity $\Sigma(\epsilon) = \sum_{n=1}^\infty \exp(- n^\epsilon)$. This sum is convergent when  $\epsilon>0$. Then set $q_n =  \exp(- n^\epsilon)/\Sigma(\epsilon)$, and note  $\sum_{n=1}^\infty q_n=1$ provided $\epsilon>0$. Then the Gibbs inequality becomes
\begin{equation}
S \leq - \sum_{n=1}^\infty p_n\, \ln q_n = \ln \Sigma(\epsilon) + \sum_{n=1}^\infty p_n \,n^\epsilon.
\label{E:gibbs2}
\end{equation}
Thus a \emph{sufficient} condition for the Shannon entropy to be finite is that there exist some $\epsilon>0$ such that
\begin{equation}
\langle n^\epsilon \rangle = \sum_{n=1}^\infty p_n \, n^\epsilon < \infty.
\end{equation}
This is of course part of a general phenomenon. 
\item
Let $E_n$ be a collection of numbers such that $Z(\beta) = \sum_{n=1}^\infty \exp(-\beta E_n)$ converges for some at least one value of $\beta$. Now define $q_n = \exp(-\beta E_n)/Z(\beta)$,  then $\sum_{n=1}^\infty q_n = 1$ provided $\beta$ is such that the sum $Z(\beta)$ converges. 
 Then the Gibbs inequality becomes
\begin{equation}
S \leq - \sum_{n=1}^\infty p_n \, \ln q_n = \ln Z(\beta) + \beta \sum_{n=1}^\infty p_n \, E_n.
\label{E:gibbs3}
\end{equation}
Thus a \emph{sufficient} condition for the Shannon entropy to be finite is that there exist some set of numbers $E_n$, and some $\beta$, such that the corresponding $Z(\beta)$ converges and such that
\begin{equation}
\langle E_n \rangle = \sum_{n=1}^\infty p_n \, E_n < \infty.
\end{equation}
\end{itemize}
On the other hand, deriving a \emph{necessary} condition requires rather different tools. Let us first \emph{re-order} (if necessary) the $p_n$ so they are in non-increasing order ($p_{n+1} \leq p_n$). Then
\begin{equation}
1 \geq \sum_{n=1}^m p_n \geq \sum_{n=1}^m p_m = m p_m.
\end{equation}
That is, with this choice of ordering, we are guaranteed $p_n \leq 1/n$. But then
\begin{equation}
S = - \sum_{n=1}^\infty p_n \ln p_n \geq  \sum_{n=1}^\infty p_n \ln n.
\end{equation}
Thus a \emph{necessary} condition for the Shannon entropy to be finite, \emph{when the probabilities are sorted into non-increasing order}, is that 
\begin{equation}
\langle \ln n \rangle = \sum_{n=1}^\infty p_n \ln n < \infty.
\end{equation}
We can eliminate the need for explicit re-ordering as follows: 
Using the previously defined object $N(p) = \sum_{n: p_n \geq p} 1$ we can define the quantities
\begin{equation}
\tilde p_n = \max\{ p: N(p)\geq n\}.
\end{equation}
Then $\tilde p_n$ is automatically a rearrangement of the $p_n$ in non-increasing order, and a \emph{necessary} condition for the Shannon entropy to be finite is that 
\begin{equation}
\langle \ln n \rangle^\sim = \sum_{n=1}^\infty \tilde p_n \ln n < \infty.
\end{equation}
The mathematical tools used so far have been extremely basic inequalities and series; the analysis has been minimal. We shall now use some slightly more sophisticated analysis in the form of Dirichlet series. 

\section{Dirichlet series}\label{S:dirichlet}

Define the generalized zeta function, a particular type of Dirichlet series~\cite{dirichlet-series}, by
\begin{equation}
\zeta_S(z) = \sum_{n=1}^\infty (p_n)^z. 
\end{equation}
One could think of the $S$ as standing either for Shannon or for entropy. A minor improvement is to explicitly exclude any states of zero probability and take
\begin{equation}
\zeta_S(z) = \sum\nolimits’_n (p_n)^z = \sum_{n: p_n>0} (p_n)^z. 
\end{equation}
By construction $\zeta_S(1)= \sum_{n=1}^\infty p_n= 1$, so this particular Dirichlet series certainly converges (absolutely) for $z\geq 1$. The interesting question is whether it converges for any $z$ less than 1. 
Note that
\begin{equation}
S = - \left.{\d \ln\zeta_S(z)\over\d z} \right|_{z=1}.
\end{equation}
If we now view $z$ as a complex number then, (in contrast to the usual situation for Taylor series where there is a \emph{radius of convergence}), for Dirichlet series there is an \emph{abscissa of convergence} $\sigma$ such that the series converges over the complex half-plane defined by~\cite{dirichlet-series}
\begin{equation}
\Re(z) > \sigma,
\end{equation}
and diverges over the complex half plane 
defined by
\begin{equation}
\Re(z) < \sigma.
\end{equation}
The line $\Re(z)=\sigma$ has to be treated delicately, in a manner somewhat analogous to the fact that that Taylor series behavior \emph{on} the radius of convergence has to be treated delicately~\cite[ see esp p 10]{dirichlet-series}. The fact that the series is clearly convergent for real $z>1$ guarantees that $\sigma \leq 1$, the abscissa of convergence is bounded above by unity. 
The relevance of this observation lies in the following fact:
\begin{equation}
\hbox{A \emph{sufficient} condition for the entropy to be finite is that $\sigma < 1$.}
\end{equation}
For a finite state space this is automatic. If we take the definition where zero probability states are excluded then the abscissa of convergence is $\sigma = -\infty$. (Even if we somewhat foolishly keep the zero probability states in the Dirichlet series, we still have $\sigma =0$.) For a countably infinite state space there is something to be proved. In particular, because all the coefficients in the generalized zeta function $\zeta_S(z)$ are positive, the real point on the abscissa of convergence is known to be a singular point of the function $\zeta_S(z)$. See~\cite[see p 10]{dirichlet-series}. The word ``singular’’ is used in the sense of ``not analytic’’, so that there is no convergent Taylor series around the point $z=\sigma$.  This happens if (for sufficiently large $m$) one of the derivatives diverges:
\begin{equation}
\zeta_S^{(m)}(\sigma) = \sum_{n=1}^\infty p_n^\sigma (\ln p_n)^m = \infty.
\end{equation}
If this happens for $m=1$ (the first derivative) then the entropy is infinite. However, this might not happen until $m>1$, perhaps even much greater than 1. That is:
\begin{equation}
\hbox{Unfortunately $\sigma<1$ is not a \emph{necessary} condition for finite entropy.}
\end{equation}

\paragraph{Example 1:}
As an explicit example of this phenomenon, recall that we had previously seen that the particular choice
\begin{equation}
p_n =  {1\over \Sigma(u) \; n \; (\ln n)^{1+u}};  \qquad \Sigma(u) = \sum_{n=\ceil{e}}^\infty  {1\over n \; (\ln n)^{1+u}};
\label{E:particular}
\end{equation}
leads to both finite entropy and normalizable probability for $u>1$. But the generalized zeta function corresponding to this particular $p_n$ is
\begin{equation}
\zeta_S(z) = \Sigma(u)^{-z} \sum_{n=\ceil{e}}^\infty {1\over n^z \; (\ln n)^{(1+u)z}}. 
\end{equation}
And for this particular zeta function it is very easy to see that the abscissa of convergence is $\sigma=1$. (See for instance related discussion in Hardy~\cite{pure}, Hardy and Riesz~\cite{dirichlet-series}, or Shilov~\cite{shilov}; the key point is that for $\Re(z)\neq1$ the $n^z$ term dominates and controls convergence/divergence of the series. For $\Re(z)=1$ one has to look carefully at the exponent of the $\ln n$ term.)  Furthermore, for this particular $p_n$ we see
\begin{equation}
\fl
\zeta_S^{(m)}(1) =  \sum_{n=\ceil{e}}^\infty p_n (\ln p_n)^m = \ln \Sigma(u) +  {1\over \Sigma(u)} \sum_{n=\ceil{e}}^\infty {(\ln n + (1+u) \ln\ln n)^m \over n (\ln n)^{(1+u)}}.\quad
\end{equation}
The dominant term in this last sum comes from the $(\ln n)^m$ in the numerator, so convergence of $\zeta_S^{(m)}(1)$ for the specific probability distribution presented in equation (\ref{E:particular}) is equivalent to convergence of 
\begin{equation}
\sum_{n=\ceil{e}}^\infty {1\over n (\ln n)^{(1+u-m)}}.
\end{equation}
But this series converges only for $u>m$. So even if the probabilities converge ($u>0$), and even if in addition the entropy converges ($u>1$), for any finite $u$ there will always be a sufficiently high derivative ($m>u$) that fails to converge.  This verifies by explicit example that $\sigma<1$ is not a \emph{necessary} condition for finite entropy. \hfill $\Box$
 
\paragraph{Example 2:}
On the other hand, let us now consider the following situation: Let $z_0>1$, and define quantities
\begin{equation}
\tilde p_n = {(p_n)^{z_0}\over \zeta_S(z_0)}.
\end{equation}
Then by construction $\sum_n \tilde p_n=1$ is absolutely convergent.  The generalized zeta function associated with $\tilde p_n$ is
\begin{equation}
\tilde \zeta_S(z) = \sum\nolimits’_n \left({(p_n)^{z_0}\over \zeta_S(z_0)}\right)^z = {\zeta_S(z_0\,z)\over \zeta_S(z_0)^z}.
\end{equation}
But this implies $\tilde \zeta_S(z)$ is convergent for $z_0 z \geq 1$, that is $z > 1/z_0$. Therefore the abscissa of convergence for the $\tilde p_n$ satisfies $\tilde \sigma \leq 1/z_0 < 1$, which implies that the $\tilde p_n$ are guaranteed to have finite Shannon entropy. Now
\begin{equation}
\ln\tilde \zeta_S(z) = \ln\zeta_S(z_0\,z) - z \ln \zeta_S(z_0).
\end{equation}
A brief computation yields
\begin{equation}
\tilde S = - \left.{\d \tilde\zeta_S(z)\over\d z}\right|_{z=1} = - z_0 \left.{\d \ln\zeta_S(z)\over\d z}\right|_{z=z_0} +\ln\zeta_S(z_0).
\end{equation}
This is certainly finite for any $z_0>1$.  So it is extremely easy to construct a very large class of probability distributions which have both finite Shannon entropy and and an abscissa of convergence strictly less than unity: $\sigma < 1$.  \hfill $\Box$

\section{Summary and Discussion}\label{S:discussion}

We have considered situations of infinite entropy, (primarily Shannon entropy, though the modifications required for dealing with von Neumann entropy are straightforward), defined over a countably finite state space.  The discussion concerns single-channel information theoretic entropy, and the additional subtleties encountered in stochastic field theories and QFT are deferred for now. We have developed a number of very simple bounds and asymptotic estimates to probe the onset of infinite Shannon entropy, with an emphasis on keeping technical computations as simple as possible. Key results are that to obtain infinite Shannon entropy an infinite number of states must have non-zero but arbitrarily small probability, that the Shannon entropy can never be too divergent, and that in a suitable technical sense infinite Shannon entropy is never worse than logarithmic in the cutoff.   The message to take from this logarithmic bound is that large Shannon entropies cannot be tightly localized in state space, large Shannon entropies must invariably come from exponentially large regions of the state space. 

\ack

VB acknowledges support by a Victoria University PhD scholarship.
MV acknowledges support via the Marsden Fund, and via a James Cook Fellowship, both administered by the Royal Society of New Zealand. 

\section*{References}

\end{document}